\documentclass[11pt,epsf]{article}

\newif\ifdohyperref
\dohyperreffalse

\dohyperreftrue

\def\circa#1{\,\raise.3ex\hbox{$#1$\kern-.75em\lower1ex\hbox{$\sim$}}\,}

\def\beq{\begin{eqnarray}}
\def\eeq{\end{eqnarray}}
\def\bea{\begin{eqnarray*}}
\def\eea{\end{eqnarray*}}

\def\centeron#1#2{{\setbox0=\hbox{#1}\setbox1=\hbox{#2}\ifdim
\wd1>\wd0\kern.5\wd1\kern-.5\wd0\fi
\copy0\kern-.5\wd0\kern-.5\wd1\copy1\ifdim\wd0>\wd1
\kern.5\wd0\kern-.5\wd1\fi}}
\def\ltap{\;\centeron{\raise.35ex\hbox{$<$}}{\lower.65ex\hbox{$\sim$}}\;}
\def\gtap{\;\centeron{\raise.35ex\hbox{$>$}}{\lower.65ex\hbox{$\sim$}}\;}
\def\gsim{\mathrel{\gtap}}

\def\singleandthirdspaced{\baselineskip=\normalbaselineskip\multiply
    \baselineskip by 130\divide\baselineskip by 100}

\newcommand{\newc}{\newcommand}
\newc{\qbar}{{\overline q}}
\newc{\Kahler}{K\"ahler }
\newc{\deltaGS}{\delta_{\rm GS}}

\ifdohyperref
  \usepackage[hyperindex]{hyperref}
  
  \newcommand\eprint[1]{\href{http://arXiv.org/abs/#1}{[arXiv:#1]}}
\else
  
  \newcommand\eprint[1]{[arXiv:#1]}
\fi

\begin{document}
\begin{titlepage}
\begin{flushright}
{\large 
hep-ph/0604140 \\ 
SCIPP-2006/05\\
SLAC-PUB-11805\\
LA-UR-06-2660
}
\end{flushright}

\vskip 1.2cm

\begin{center}

{\LARGE\bf Moduli Decays and Gravitinos}

\vskip 1.4cm

{\large Michael Dine$^a$, Ryuichiro Kitano$^b$, Alexander Morisse$^a$,  
Yuri Shirman$^c$}
\\
\vskip 0.4cm

{\it $^a$Santa Cruz Institute for Particle Physics,
     Santa Cruz CA 95064  } \\
{\it $^b$Stanford Linear Accelerator Center,
     Stanford University, Stanford, CA 94309} \\
{\it $^c$T-8, MS B285, LANL, Los Alamos, NM 87545 }\\

\vskip 4pt

\vskip 1.5cm

\begin{abstract}
  One proposed solution of the moduli problem of string cosmology
  requires that the moduli are quite heavy, their decays reheating the
  universe to temperatures above the scale of nucleosynthesis.  In
  many of these scenarios, the moduli are approximately
  supersymmetric; it is then crucial that the decays to gravitinos are
  helicity suppressed.  In this paper, we discuss situations where
  these decays are, and are not, suppressed.  We also comment on
  a possible gravitino problem from inflaton decay.
\end{abstract}

\end{center}

\vskip 1.0 cm

\end{titlepage}
\setcounter{footnote}{0} \setcounter{page}{2}
\setcounter{section}{0} \setcounter{subsection}{0}
\setcounter{subsubsection}{0}

\singleandthirdspaced

\section{Introduction}

It is possible that nature exhibits an approximate $N=1$
supersymmetry.  In string theory, such an approximate supersymmetry is
often accompanied by approximate moduli.  For cosmology, these moduli
are both intriguing and problematic.  Intriguing because it is
tempting to connect them with inflation; problematic because they tend
to carry too much energy, spoiling the successes of big bang
nucleosynthesis.  One possible resolution of these problems is to
suppose that the moduli have masses well above the masses of squarks
and gauginos
(assumed to be of order TeV or so)~\cite{bkn}.  Then the moduli decay
reheats the universe to temperatures above nucleosynthesis
temperature.  In such a scenario, the main issues are production of
dark matter and gravitinos.  Production of dark matter was addressed
in an early paper of Moroi and Randall~\cite{moroirandall}.  These
authors argued that decays of moduli to gauginos were helicity
suppressed, and used this in a series of estimates of the dark matter
density.  
It has been argued that similar suppression ratio takes place
for the branching ratio to gravitinos\footnote{In reference
\cite{moroirandall}, this question was moot since the moduli masses
were assumed to be of order $m_{3/2}$.}~\cite{moroirandall, Hashimoto:1998mu}.
Subsequently, other authors have considered variations of this scenario,
supposing similar helicity suppressions~\cite{kohrietal}.

Moroi and Randall~\cite{moroirandall} considered situations where the
moduli masses were 
comparable to $m_{3/2}$.  But it is also possible that the moduli have
masses which are large compared to $m_{3/2}$, and approximately
supersymmetric.  This is the situation in the model of Kachru, Kallosh,
Linde and Trivedi (KKLT)~\cite{kklt}, and it has been assumed by many
authors that there is a similar suppression there.  Recently, however,
two groups have examined this question. 
In \cite{endo} and \cite{yamaguchi} it was
argued, quite generally, that
neither the gravitino nor the gaugino branching ratio are helicity
suppressed.  Both sets of authors argue specifically that in the models
of~\cite{kklt}, the decay rate is unsuppressed.

In this note, we look at both of these questions.  We find, in agreement
with both groups, that in models with
supersymmetric moduli, there is, in general, no suppression of the
gaugino decay rate.  The situation of the gravitino is more complicated.
We will exhibit examples where there is no suppression of the
gravitino rate.  
But in models like that of KKLT, with a simple supersymmetry breaking
sector, we show that there is a suppression.
As guidance for our analysis, we rely heavily on the
Goldstino equivalence theorem.  This theorem is analogous to the
equivalence theorem in spontaneously broken gauge theories, which
asserts that at high energies the amplitudes for processes involving
longitudinal gauge bosons are the same as those for the would-be
Goldstone bosons of the symmetry breaking.  In supergravity, one just
replaces ``gauge bosons" by ``gravitinos" and ``Goldstone bosons" by
``Goldstinos" to obtain the corresponding equivalence theorem.  In the
KKLT-type models, working directly with the Goldstinos, it is easy to
see that all of the decay amplitudes are suppressed by $m_{3/2}$.  With
a bit more effort, one can find the suppression in terms of the
longitudinal gravitinos of unitary gauge.  For the models without
suppression, it is easy to see the equivalent descriptions.

We will see, though, that whether or not there is suppression depends
strongly on the assumptions about supersymmetry breaking.  If the hidden
sector contains approximate moduli with mass of order $m_{3/2}$, the
decays of the heavy moduli to gravitinos are suppressed; if not, they
are unsuppressed.  This observation resuscitates the moduli problem,
since one either suffers from a gravitino problem or another modulus.
On the other hand, in models like KKLT, the gravitino mass is expected
to be quite large, so the decays of this other modulus can themselves
reheat the universe to nucleosynthesis temperatures.  
So, with a slight modification of the original scenario
\cite{moroirandall} of Moroi and
Randall, a viable cosmology is possible. 

Having established criteria for suppression, we briefly discuss the
implications of these observations for various cosmological models in
the literature, especially that of KKLT. We also comment on the
inflaton decay into gravitinos, which is discussed recently in
Ref.~\cite{Kawasaki:2006gs}.  More detailed studies will appear in a
subsequent publication.

\section{Some Studies of the Goldstino Equivalence Theorem}
\label{sec:equivalence}

The validity of the equivalence theorem follows from a simple physical
argument.  Consider a theory with supersymmetry broken at a low energy
scale, $M_{susy}^2=F$, well below the {\it intermediate scale},
$M_{int}^2= {\rm TeV} \cdot M_p$.  
For the Goldstone particles of
this theory, and their would-be superparners, gravity is completely
irrelevant.  For example, in a hidden-sector-type model, approximate
moduli will interact with gravitinos with interaction strength $1/F
\gg 1/M_p^2$.

As a simple model, consider a theory with a massive field, $\phi$, and a
Polonyi sector, with fields $z$, and superpotential 
\beq
W = W_o +  \mu^2 z + {M \over 2} \phi^2 \,
\eeq
where
$W_o$ is adjusted so that the cosmological constant
vanishes (this is helpful conceptually
but  will not be particularly important for our
discussion).
In this model, supersymmetry is broken.  
The Goldstino is the fermionic component of $z$, $\psi_z$.
We take a \Kahler potential including interaction between $\phi$ and $z$
supermultiplets: 
\beq 
K =
\phi^\dagger \phi + z^\dagger z + {1 \over \Lambda}\phi^\dagger z^2 +
{\rm c.c.}\ .  
\eeq 
$\Lambda$ might be the mass of some fields which have been integrated
out.
With this interaction term, the minimum of the potential lies at $z \sim
\Lambda^2 / M_p$.
The modulus coupling to gravitinos becomes
\beq {1  \over \Lambda} F_\phi^\dagger \psi_z \psi_z 
= {M \over \Lambda} \phi \psi_z \psi_z \,.  
\eeq 
This leads to an amplitude: 
\beq {\cal A}(\phi
\rightarrow \psi_z \psi_z) = {M \over \Lambda}.
\label{toyamplitude}
\eeq

We can find the same result in terms of the longitudinal gravitinos.
The basic coupling, which will figure repeatedly in our discussions,
is:
\beq
{\cal L}_{cv} = -
e^{G/2} \left[
\psi_\mu \sigma^{\mu \nu} \psi_\nu + {\rm h.c.}\right].
\eeq
Here $G = K + \ln(W) + \ln(W^*)$ 
and we set $M_p=1$ 
whenever this is unambiguous.  
The coupling to scalars, $\phi_i$ arises from Taylor series
expanding the exponential:
\beq
\label{eq:u-coupling}
{\cal L}_{cv}= - {e^{G/2} \over 2}\ 
G_i \phi_i \psi_\mu \sigma^{\mu \nu} \psi_\nu.
\eeq
In our case, $G_\phi =0$, $e^{G/2} G_z = \mu^2$.  
The massive field is an admixture of $\phi$ and $z$.
Eliminating $F_\phi$ by its
equation of motion, the mass matrix has the structure:
\beq
M^2 \vert \phi \vert^2 + {2 \over \Lambda} 
F_z M \phi z
+ {\rm c.c.}
\eeq
up to terms of ${\cal O} (m_{3/2}^2)$. 
As a result, the massive scalar is:
\beq
\Phi = \phi + {2 F_z \over M \Lambda} z^\dagger \ .
\eeq

The longitudinal gravitino is:
\beq
\psi_\mu(k) = \sqrt{2 \over 3} {k_\mu \over m_{3/2}} u_\alpha(k).
\eeq
This can be obtained from the Rarita-Schwinger action, or by writing the
supersymmetry transformation to unitary gauge (the $\sqrt{3}$ factor results
from the relation $\vert G_z \vert^2 =3$ which holds for vanishing
cosmological constant).  
Using the Dirac equation, and keeping only terms of order $k/m_{3/2}$ gives
for the amplitude:
\beq
{\cal A} (\Phi \rightarrow \psi_\mu \psi_\mu)
= {2 \over 3}\left ({k \cdot k^\prime \over m_{3/2}^2}
\right ) {e^{G/2} \over 2} {2 F_z \over M \Lambda} \ G_z
.
\eeq
Using $F_z^2 = 3 m_{3/2}^2$,
$e^{G/2} = m_{3/2}$
and 
$2 k \cdot k^\prime =M^2$
gives precisely the amplitude of eqn.~(\ref{toyamplitude}).

Note that this amplitude is not chirality suppressed.  If we take
$\Lambda=M_p$, this is of precisely the size found, in another
context, in \cite{endo,yamaguchi}.

\section{The KKLT Model}
\label{sec:kklt}

In the simplest version of the KKLT model, the only moduli which are light
relative to the fundamental scale are the \Kahler moduli.
We will assume that there is only one such modulus, which we will call
$\rho$.  
Supersymmetry breaking can arise, as suggested by KKLT, due to the
presence of anti-D3 branes. Alternatively, similar scaling for the
moduli potential may result from the hidden sector
fields~\cite{endo, Lebedev:2006qq}. We will first consider the simplest
hidden sector model, with a single field, $z$.
We will take the K{\" a}hler potential and
superpotential to be
\beq
K = -3 \ln(\rho + \rho^\dagger) + z^\dagger z\ ,~~~~
W = e^{-\rho} + W_o + \mu^2 z.
\eeq
Here we have taken various coefficients to be numbers of order unity.
(For simplicity, we have taken
the coefficient of $\rho$ in the exponential to be $-1$; our
analysis does not depend on this choice.)
$W_o$ is assumed to be very small.  
The supersymmetry breaking effect in the hidden sector ($z$ sector) is
mediated to the $\rho$ field through gravitational ($1/M_p$ suppressed)
interaction.
To leading order in $1/\rho$,  the $\rho$ sector is
supersymmetric and the vacuum is determined by
\beq
D_\rho W = 0\ , ~~~~~  V_z =0.
\eeq
Here $D_\rho$ denotes the \Kahler derivative,
\beq
D_\rho W = {\partial W \over \partial \rho} + 
{\partial K \over \partial \rho}W
= W G_\rho.
\eeq
In this case,
\beq
z \approx \sqrt{3}-1\ ,~~~~
W_o \approx (2 - \sqrt{3})\mu^2\ , ~~~~
\rho \approx -\ln(W_o/\ln(W_o)) \gg 1\ ,
\eeq
and $m_{3/2} = {\cal O}(\mu^2)$.
The $\rho$ field has a supersymmetric mass of order $\rho m_{3/2} \gg
m_{3/2}$ in this vacuum, which justifies the approximation $G_\rho
\approx 0$.
Gravity-mediated supersymmetry breaking effects give corrections to
the $\rho$ potential characterized by $1/\rho$.  In counting powers of
$\rho$, it is useful to note that $W_\rho \sim W/\rho $, and
$W_{\rho \rho} \sim W_\rho$.

One can now study the amplitude for the modulus decay in powers of
$1/\rho$. 
In the unitary gauge the amplitude is proportional to $G_\rho$ (see
eqn.~(\ref{eq:u-coupling})).
As argued in \cite{endo, yamaguchi}, in a model of this
type, $G_\rho$ would receive corrections of order $1/\rho$.  Recalling
the structure of the unitary gauge amplitude, and that $m_\rho \sim
\rho m_{3/2}$, it would seem that the amplitude is enhanced over naive
expectations by a factor of $m_\rho/m_{3/2}$.

It is difficult, however, to find such an enhancement in the
Goldstino picture.  Here there are various types of terms.  First,
there are terms in the action involving derivatives.  However, all
of these {\it do have a chirality suppression}.  Then there are helicity-flip
terms, involving two matter fermions.  
The relevant terms in the supergravity action are:
\beq
{\cal L}_{\chi \chi} = - {1 \over 2}e^{G/2}
\left(  G_{ij} + {1 \over 3} G_i G_j - \Gamma_{ij}^k G_k \right)
\psi_i \psi_j,
\label{eq:bilinear}
\eeq
where the connection is defined by $\Gamma_{ij}^k = g^{k \bar l} G_{ij
\bar l}$.
To obtain the coupling of the heavy modulus to Goldstinos we need to
Taylor expand the factors of $G$ in powers of the
modulus and project matter fermions onto Goldstinos. 
The Goldstino is a linear combination
of $\psi_\rho$ and $\psi_z$, of the form:
\beq
\tilde G = \cos{\theta}\psi_z + \sin{\theta} \psi_\rho \,,
\label{eq:a-factor}
\eeq
where $\theta \sim 1/\rho$, as we will see shortly.  For the calculation
of the leading contribution in the $1/\rho$ expansion, we can take
$\psi_i=\psi_j=\psi_z$ which gives a coupling of $\rho$ to a pair of
Goldstinos:
\beq
-{1 \over 2}
e^{G/2} 
\left(
G_{zz\rho} + {2 \over 3} G_{z\rho} G_{z}
\right) \rho \psi_z \psi_z.
\eeq
But both terms in parenthesis are of order $1 / \rho$.  When we rescale $\rho$
to obtain a canonical kinetic term, we obtain an amplitude of order $m_{3/2}$.
We obtain the same result if we take one of the fermions to be $\psi_\rho$; the
projection onto the Goldstino again leads to a term of order $m_{3/2}$.

How does one reconcile the results of these two different gauges
(descriptions)?  The physical Goldstino is a linear combination of
$\psi_z$ and $\psi_\rho$; the orthogonal combination, $\Psi$, is
massive. Similarly, the heavy scalar $\Phi$ is a linear combination of
$z$ and $\rho$. In fact, it is precisely this mixing that is responsible
for the non-vanishing $G_\rho$
found in Refs.~\cite{endo, yamaguchi}.
However, in the KKLT model under consideration the mixing is
supersymmetric at order $1/\rho$! This means that at this order the
only non-vanishing $F$-term is in the same supermultiplet as the
true Goldstino.
As we will now explain, for the heavy field
\beq
G_\Phi = {\cal O} \left({1 \over \rho^2} \right).
\eeq

It is very helpful to work with the
supergravity action written in terms of the quantity $G$, rather than
$K$ and $W$ separately.  In terms of $G$, the potential is: 
\beq V = e^G
[g^{i \bar j} G_i G_{\bar j} -3].  
\eeq 
There are two things we need to calculate: the $F$ term for $\rho$ --
which is now the problem of finding $G_\rho$, and the eigenstates of the
mass matrix.
From the condition that $\partial V / \partial \rho^\dagger = 0$, we find
\begin{eqnarray}
 G_\rho = - 
\left[
\frac{ \rho + \rho^\dagger }{ \sqrt{3}}
\right]^{-2}
\frac{ G_{\bar \rho \bar z} }{G_{\bar \rho \bar \rho}} 
\ G_{z}\ ,
\end{eqnarray}
at the leading order in the $1/\rho$ expansion. Indeed this is of ${\cal
O}(1/\rho)$ after the field rescaling such that $\rho$ has the
canonically normalized kinetic term, i.e., $\rho \to (\langle \rho +
\rho^\dagger \rangle / \sqrt{3} ) \rho$ and thus $G_\rho \to (\langle
\rho + \rho^\dagger \rangle / \sqrt{3}) G_\rho $.

Now let's consider the mass matrix. We will keep the leading order
contribution in the $1/\rho$ expansion for each component. After the
rescaling of $\rho$ we have 
\begin{eqnarray}
 V_{\rho \bar z} = e^G \left[
\frac{\langle \rho + \rho^\dagger \rangle }{\sqrt{3}}
\right]^{3} G_{\rho \rho} G_{\bar \rho \bar z}\ ,\ \ \ 
 V_{\rho \bar \rho} = e^G \left[
\frac{\langle \rho + \rho^\dagger \rangle }{\sqrt{3}}
\right]^{4} G_{\rho \rho} G_{\bar \rho \bar \rho}\ ,
\end{eqnarray}
and $V_{z \bar z}$ is of the order of $e^G$. Note that these terms are
supersymmetric; there are corresponding large terms in the fermion mass
matrix.
Because of this mixing, the mass eigenstate of the heavy scalar has a
small component of the field $z$, such that
\begin{eqnarray}
 \Phi = \hat{\rho} + 
\epsilon^* z\ ,
\end{eqnarray}
where 
\begin{eqnarray}
 \epsilon = \left[
\frac{\langle \rho + \rho^\dagger \rangle }{\sqrt{3}}
\right]^{-1} \frac{G_{\bar \rho \bar z}}{G_{\bar \rho \bar \rho}}\ ,
\label{eq:epsilon}
\end{eqnarray}
and $\hat{\rho}$ represents the canonically normalized $\rho$ field.
The fermion has the same mixing factor, and thus the angle $\theta$ in
eqn.~(\ref{eq:a-factor}) is given by $\theta=-\epsilon$.
The $G_\Phi$ factor is calculated to be
\begin{eqnarray}
 G_\Phi = \epsilon G_z 
+ \frac{\langle \rho + \rho^\dagger \rangle }{\sqrt{3}} G_\rho = 0\ .
\end{eqnarray}
The ${\cal O}(1/\rho)$ contribution cancels out and the unitary gauge
and Goldstino picture are reconciled.
At next order, there is no reason for things to vanish.  But this corresponds
to amplitudes which are suppressed by two powers of $\rho$,
i.e. which are proportional to $m_{3/2}$.

We should note, however, that the suppression we have found here does not
result from general symmetry principles, but is a feature of this particular
model.  
We can modify the KKLT \Kahler potential in a way which yields a decay
amplitude as large as that suggested 
in \cite{endo, yamaguchi}. If we take 
\beq
K = -3 \ln(\rho + \rho^\dagger - z^\dagger z + \kappa (zz + z^\dagger
z^\dagger) ), 
\eeq
where $\kappa$ is a real parameter,
then the Christoffel connection $\Gamma_{zz}^\rho$ has a non-vanishing value
$\Gamma_{zz}^\rho = 2 \kappa$.
In this case, the terms in the supergravity Lagrangian:
\beq
{1 \over 2} e^{G/2} \Gamma_{ij}^k  G_k \psi^i \psi^j
= \kappa m_{3/2} G_\rho \psi_z \psi_z + \cdots
\eeq
give a coupling of order
\beq
\kappa m_{3/2} G_{\rho \rho} \rho \psi_z \psi_z \simeq
{-3 \over \langle \rho + \rho^\dagger \rangle} \kappa m_{3/2} \rho \psi_z \psi_z.
\eeq
Note that this term vanishes when $\kappa =0$, as in the KKLT
model in Ref.~\cite{endo}. The other terms are ${\cal O}(1/\rho^2)$.
Rescaling the $\rho$ and $z$ fields so that their kinetic terms
are canonical, this gives a decay amplitude of order $\rho m_{3/2}$,
i.e. of order $m_\rho$.  
So determining the presence or absence of chirality
suppression requires detailed microphysical understanding of the model.

In general, for a particle $\phi$ which has a supersymmetric mass
$m_\phi \gg m_{3/2}$, only the third term in the parenthesis in
eqn.~(\ref{eq:bilinear}) has the potential of enhancing the decay
amplitude.  With the canonically normalized kinetic terms for $\phi$ and
$z$, the Taylor expansion of the term gives
\begin{eqnarray}
{\cal L} &\ni& {1 \over 2} e^{G/2}
\Gamma_{zz}^\phi G_{\phi \phi} \phi \psi_z \psi_z +
{\rm h.c.} \nonumber \\
&\sim& m_\phi \Gamma_{zz}^\phi \phi 
\psi_z \psi_z +
{\rm h.c.}
\end{eqnarray}
Here we have used $G_{\phi \phi} \sim W_{\phi \phi} / W \sim m_\phi /
m_{3/2}$.  Therefore there is a helicity suppression unless a direct
$\phi^\dagger z z$ coupling is present in the \Kahler potential such
that $\Gamma_{zz}^\phi = {\cal O}(1)$.

\section{Models Without Light Moduli in the Hidden Sector}
\label{sec:light-moduli}

It was crucial to the cancellation of the previous section that 
the mixing between $\rho$ and $z$ was supersymmetric, and in
particular that all soft masses were small compared to the mass of the
heavy modulus.
One might suspect that if
the 
soft mass of the field $z$ was {\it large} compared to $m_\rho$, then
this cancellation would no longer occur.  Such a large mass might arise
if supersymmetry is broken dynamically~\cite{Affleck:1983rr,
Affleck:1983mk, Affleck:1984xz}, or if the hidden sector involves
non-trivial interactions, as in the O'Raifeartaigh model.  To show that
there is no cancellation, in general, we can modify the \Kahler
potential for the field $z$ so that it obtains a large,
non-supersymmetric mass: \beq \delta K = - {1 \over \Lambda^2}
(z^\dagger z)^2.  \eeq Here $\Lambda \ll M_p$.
This model effectively describes the situation without flat directions
such as models with dynamical supersymmetry breaking, e.g., in
Ref.~\cite{Affleck:1983rr, Affleck:1983mk, Affleck:1984xz}. In general,
the above term can be generated by integrating out heavy fields with
masses $\Lambda \gsim (m_{3/2} M_p)^{1/2}$.
In this case, there is no longer a cancellation, in the
unitary gauge picture. The above term gives a mass term for the $z$ field
\begin{eqnarray}
 m_{\rm Hidden}^2 \simeq 12 m_{3/2}^2 
\left( \frac{M_p}{\Lambda} \right)^2
\gg m_{3/2}^2\ .
\label{eq:lambda-mass}
\end{eqnarray}
When $z$ is much heavier than $\rho$ in the KKLT example, the
lighter mass eigenstate $\Phi$, which mainly consists of $\rho$, is
obtained to be
\begin{eqnarray}
 \Phi = \hat{\rho} + \epsilon^{\prime *} z\ ,
\end{eqnarray}
where $\epsilon^\prime$ is\footnote{Note that this expression is only
  valid for $m_{\mathrm{Hidden}}\circa{>} m_{\rho}$, otherwise the
  mixing factor is given by $\epsilon$ in eqn.~(\ref{eq:epsilon}) with
  ${\cal O}(m_{\rm Hidden}^2/\rho m_\rho^2)$ corrections.}
\begin{eqnarray}
 \epsilon^\prime = {\cal O} \left(
{1 \over \rho} {m_\rho^2 \over m_{\rm Hidden}^2}
\right)\ .
\end{eqnarray}
Therefore, for $m_\rho \ll m_{\rm Hidden}$,
\begin{eqnarray}
 G_\Phi = 
\frac{\langle \rho + \rho^\dagger \rangle }{\sqrt{3}} G_\rho 
+ \epsilon^\prime G_z 
\simeq \frac{\langle \rho + \rho^\dagger \rangle}{\sqrt{3}} G_\rho
= {\cal O} \left( {1 \over \rho} \right)\ .
\end{eqnarray}
The cancellation does not take place in this case. The amplitude is
${\cal O}(m_\rho)$.

In the Goldstino picture, it is the $\Gamma^{z}_{zz}$ coupling
which leads to the unsuppressed decay.
The term in the Lagrangian,
\begin{eqnarray}
{1 \over 2} e^{G/2} \Gamma_{zz}^z G_z \psi_z \psi_z
\end{eqnarray}
contains a coupling of $\Phi$ to Goldstinos through the mixing $\epsilon^\prime$:
\begin{eqnarray}
{\epsilon^{\prime *} \over 2 } m_{3/2} G_{zz \bar z \bar z} G_z
 \Phi^\dagger \psi_z \psi_z\ .
\end{eqnarray}
With $G_{zz\bar z \bar z} = - 4/\Lambda^2$ and $G_z = \sqrt{3}$, the
amplitude is estimated to be ${\cal O}(m_\rho)$, consistent with the
calculation in the unitary gauge.\footnote{ The situation with
explicitly broken supersymmetry can be realized in the limit of $m_{\rm
Hidden} \to \infty$, although the description in terms of effective
field theory would break down above the intermediate scale $(m_{3/2}
M_p)^{1/2}$. This is analogous to the ``Higgsless'' limit ($m_{\rm
Higgs} \to \infty$) of the electroweak symmetry breaking where the
unitarity of the scattering amplitudes of the $W$ bosons is violated at
high energy.
Ultraviolet completion of such theories may be possible with strongly
coupled gauge theories (dynamical symmetry breaking) or a warped
extra-dimension (``Higgsless model''~\cite{Csaki:2003dt}) which
necessarily imply extra states at $m_{\rm Hidden}\sim(m_{3/2}
M_p)^{1/2}$ and recovers the picture of spontaneous SUSY breaking.  }

This leaves us with a cosmological problem.  Either $z$ is very heavy,
in which case $\rho$ decays to gravitinos are problematic.  Or $z$ is
light compared to $\rho$, in which case the modulus $z$ is potentially
problematic.

But in a scenario like that of KKLT, $z$ might behave in precisely the
manner envisioned by Moroi and Randall~\cite{moroirandall}.\footnote{
The possibility of evading the gravitino problem by the presence of
moduli in the hidden sector is noted in Ref.~\cite{Lebedev:2006qq}.}
The gravitino in such a model is likely to be quite massive, easily
$100$ TeV or so~\cite{Choi:2004sx, Choi:2005ge, Endo:2005uy}.
Similarly, then, the $z$ field will be very massive.  If its mass is
less than twice the gravitino mass, it will not decay to gravitinos.
In this circumstance, an acceptable cosmology is possible with a slight
modification of the scenario of Moroi and Randall, even if $z$ dominates
the energy density of the universe.
A large enough reheating temperature for nucleosynthesis can be obtained
from $z$ decay by introducing an explicit coupling of $z$ to vector-like
visible
sector particles ($X$ and $\bar{X}$) in the \Kahler potential
\begin{eqnarray}
 \lambda {1 \over M_p} z^\dagger \bar{X} X + {\rm h.c.}.
\end{eqnarray}
If the coupling constant, $\lambda$, is ${\cal
O}(1)$, the reheating temperature is above nucleosynthesis temperatures.
Moroi and Randall discussed this scenario with the role of $X$ and
$\bar{X}$ being played by the Higgs fields, but in our case the $X$ and
$\bar{X}$ fields cannot be particles in the minimal supersymmetric
standard model (MSSM) since the above term gives a mass term of ${\cal
O}(100~{\rm TeV})$.
If, instead, we introduce $X$ and $\bar{X}$ as new fields and assign them
the
same quantum numbers and $R$-parity as the Higgs fields,
they can decay quickly through mixing with Higgs to ordinary quarks
and leptons. The
production of the $R$-parity odd particles is suppressed compared to the
quark/leptons or radiation and, therefore, the overproduction of the
dark matter can be avoided.

Alternatively, we can arrange $m_{\rm Hidden}$ such that $m_{\rm Hidden}
< m_\rho$ but $z$ decays earlier than $\rho$. However, gravitino
production in $z$ decay is potentially problematic, even accounting
for the
dilution effects from the $\rho$ decay; precisely how serious this
problem is in this model depends on the details
of the cosmological history, such as the initial amplitude of the $z$ and
$\rho$ oscillations.

\section{Gaugino Emission}

In Refs.~\cite{endo, yamaguchi}, it is also pointed out that the decay
of heavy moduli to gauginos is not necessarily helicity suppressed.
This is easy to see.  Consider a modulus $S$ with a
coupling to
gauge fields through
\beq
{\cal L}_{gauge} = f(S) W_\alpha^2 \, ,
\eeq
and the \Kahler potential
\beq
-3\log(S+S^\dagger) \,.
\eeq
Correspondingly, there is a coupling to a pair of canonically
normalized gauginos,
\beq
{\cal L}_{\lambda \lambda} = \frac{f^\prime(S)}{f(S)} g^{S\bar S}
e^{G/2}G_{\bar S} 
\lambda \lambda. 
\eeq
Then writing $S = S_o + \hat S$, where $\hat S$ represents the
fluctuating 
field, the coupling of $\hat S$ (after rescaling it to canonical
normalization) to $\lambda \lambda$ is
\beq
\label{eq:s-coupling}
g^{S\bar S} e^{G/2} G_{\bar S \bar S} \hat S \lambda \lambda 
\approx M_S \hat S \lambda \lambda 
\eeq
where $M_S$ is the supersymmetric part of the $S$ mass.  In the above
we assumed a generic scaling $f^\prime(S)/f(S) \sim 1/S$. Although
(\ref{eq:s-coupling}) is a consequence of the specific form
of the \Kahler potential, this result is quite general.

Provided
that $M_S$ is large compared to the gravitino or
gaugino mass, there is
clearly no suppression by powers of $m_\lambda$.  The branching ratio
of $S$ to gauge bosons and gauginos will be of order one.  
We should note that, unlike the possible enhancement of the gravitino rate,
which is a cosmological catastrophe, the enhancement of the gaugino
rate, for a very heavy scalar decay, is not necessarily a problem.  For
a modulus with mass of order $1000$ TeV or so, the annihilation rate for
gauginos produced in decays is large enough that one can naturally obtain
a density suitable for the dark matter density.

\section{Gravitinos from Inflaton Decays}

In Ref.~\cite{Kawasaki:2006gs}, it was argued
that there is a gravitino problem from inflaton decays. The authors 
assumed that there is no helicity suppression for
inflaton decay into gravitinos, and obtained 
quite stringent constraints 
from nucleosynthesis or overproduction of the dark
matter. 
As a result, they argued that it is critical to a successful cosmology
that $G_\phi=0$ at the minimum of the inflaton potential.\footnote{
Cosmology with a general heavy scalar field has also been studied very
recently in Ref.~\cite{asakaetal} .  }
However, in light of our previous discussion, it is natural to ask
whether inflaton decays {\it can} be helicity suppressed.

As we have already seen, the question
of whether these helicity suppressed or not depends crucially on
the structure of the hidden sector.  The decay is helicity suppressed
unless there is a direct coupling to the hidden sector of a specific type,
or the hidden sector field with non-vanishing $F$-term is heavier than the
inflaton.

We demonstrate this helicity suppression by analyzing a
very simple model with
\begin{eqnarray}
 K = \phi^\dagger \phi + z^\dagger z 
- \frac{(z^\dagger z)^2}{\Lambda^2}\ ,
\end{eqnarray}
\begin{eqnarray}
 W = \frac{m_\phi}{2} (\phi - \phi_0)^2 + \mu^2 z + W_o\ .
\end{eqnarray}
This is essentially the same situation as that of the models discussed in
\cite{Kawasaki:2006gs} and of the KKLT model analyzed in sections~\ref{sec:kklt}
and~\ref{sec:light-moduli}. The inflaton field has a large
supersymmetric mass term $m_\phi$ compared to the gravitino and the
minimum of the potential $\phi = \phi_0$ is displaced from the origin. We have again
introduced the $-(z^\dagger z)^2 / \Lambda^2$ term such that we can treat
the $z$ mass, $m_{\rm Hidden} = {\cal O} ( m_{3/2} M_p / \Lambda )$, as a free parameter.

When $m_\phi > m_{\rm Hidden}$, the decay amplitude is suppressed by a
factor of $m_{3/2}/m_\phi$ or $(m_{\rm Hidden}/m_\phi)^2$ as we see in
the following.  In the unitary gauge calculation, the mismatch between
the direction of the $F$-term and the inflaton mass eigenstate appears
to be small compared to the naive estimation of ${\cal
O}(m_{3/2}/m_\phi)$ by a factor of ${\cal O}(m_{\rm Hidden}^2/m_\phi^2)$
or ${\cal O} (m_{3/2}/m_\phi)$.
Explicitly,
the $G_\phi$ and $G_z$ factors are obtained to be
\begin{eqnarray}
 G_\phi = 
- \frac{G_{\bar z \bar \phi}}{G_{\bar \phi \bar \phi}} G_z
= - \frac{\sqrt{3} \phi_0 m_{3/2}}{m_\phi} G_z\ , \ \ \ 
G_z = \sqrt{3}
\end{eqnarray}
at the leading order in the $m_{3/2}/m_\phi$ expansion. The ${\cal
O}(m_{3/2}/m_\phi)$ contribution appeared in $G_\phi$ with a factor $\phi_0$
representing the displacement from the origin.
On the other hand,
the inflaton mass eigenstate $\Phi$ has a small $z$ component
\begin{eqnarray}
\Phi = \phi + \epsilon^{\prime \prime *} z\ ,
\end{eqnarray}
with 
\begin{eqnarray}
 \epsilon^{\prime \prime} = \frac{\sqrt{3} \phi_0 m_{3/2}}{m_\phi} 
+ {\cal O} \left( \frac{m_{3/2}}{m_\phi} \frac{m_{\rm Hidden}^2}{m_\phi^2},
\frac{m_{3/2}^2}{m_\phi^2}
\right).
\end{eqnarray}
Therefore, the $G_\Phi$ factor is
\begin{eqnarray}
 G_\Phi = \epsilon^{\prime \prime} G_z + G_\phi
=
{\cal O} \left( \frac{m_{3/2}}{m_\phi} \frac{m_{\rm Hidden}^2}{m_\phi^2},
\frac{m_{3/2}^2}{m_\phi^2}
\right)\ .
\end{eqnarray}
The ${\cal O}(m_{3/2}/m_\phi)$ contribution cancels as anticipated.
Therefore, the decay amplitude has either $(m_{\rm Hidden}/m_\phi)^2$ or
$m_{3/2}/m_\phi$ suppression.
The same result can be obtained in the Goldstino picture. The
$\Gamma_{zz}^z$ term and the mixing gives the coupling of the
inflaton to Goldstinos:
\begin{eqnarray}
{\epsilon^{\prime \prime *} \over 2 } m_{3/2} G_{zz \bar z \bar z} G_z
 \Phi^\dagger \psi_z \psi_z
\simeq
{1 \over 2} m_{\phi}  \phi_0
\left(
m_{\rm Hidden} \over m_\phi
\right)^2
\Phi^\dagger \psi_z \psi_z\ .
\end{eqnarray}
On top of the factor $\phi_0$, the amplitude has a suppression factor of
$(m_{\rm Hidden}/m_\phi)^2$.
Other terms in the Lagrangian give ${\cal O}(m_{3/2}/m_\phi)$ suppressed
amplitudes.

With sufficiently small $(m_{\rm Hidden}/m_\phi)^2$ and $m_{3/2}/m_\phi$, we
can easily satisfy the constraint from gravitino
cosmology~\cite{Pagels:1981ke, moroilimits}. The model predictions are obtained by
multiplying the larger of two factors $m_{\rm Hidden}^2/m_\phi^2$ or
$m_{3/2}/m_\phi$ by the value of $G_\phi$ presented in Figure~1 in
Ref.~\cite{Kawasaki:2006gs}.
If we are to avoid the gravitino production from the $z$ field, the
decay of $z$ should happen earlier than the inflaton decay. This gives a
constraint on the reheating temperature of the universe $T_R$. The naive
expectation for the decay width of $z$ is
\begin{eqnarray}
 \Gamma_z \sim \frac{1}{4 \pi} \frac{m_{\rm Hidden}^3}{\Lambda^2}\ ,
\end{eqnarray}
and $\Lambda^2$ is related to $m_{\rm Hidden}$ by
eqn.~(\ref{eq:lambda-mass}). Indeed the partial decay width into the
gravitinos is this size.  By comparing the inflaton decay width
$\Gamma_\phi \sim T_R^2 / M_p$, we obtain a condition
\begin{eqnarray}
 T_R \ll 10^9 ~{\rm GeV}\ 
\left(
{m_{3/2} \over 100~{\rm GeV}}
\right)^{-1}
\left(
{m_{\rm Hidden} \over 10^8~{\rm GeV}}
\right)^{5/2}\ .
\end{eqnarray}
Consistent parameter regions can be easily found.

\section{Conclusions:  Cosmological Implications}

We began this paper with studies of the equivalence theorem for gravitinos.
In several exercises, we saw that it is usually easy to compute amplitudes
for particle decays to gravitinos in a Goldstino picture, but that in unitary
gauge there are sometimes subtle cancellations, and it is easy to
overestimate these amplitudes.  From these examples, we learned that,
while there is not necessarily a helicity suppression
in the decays of
massive scalars to Goldstinos, there often is.  Perhaps of greatest
current interest, we saw that there is a suppression by $m_{3/2}$ in the simplest
version of the KKLT model.  On the other hand,
we saw that if we alter the form of the \Kahler potential in a specific
way, we can obtain an enhanced result. 

Working in the Goldstino picture, we can easily identify the conditions
for the enhancement of the decay amplitude; presence of the direct
coupling between moduli to the hidden sector field ($\phi^\dagger z z$)
in the \Kahler potential or a large supersymmetry breaking mass term for
the scalar partner of the Goldstino (e.g., through $(z^\dagger z)^2 /
\Lambda^2$ term in the \Kahler potential).
Both enhancements can be understood by looking at the coupling of the
Goldstino bilinear to the scalar field through the Christoffel
connection (the third term in eqn.(\ref{eq:bilinear})).
Whether the amplitude is suppressed or not turns out highly model
dependent.

Even when there is a suppression, however, the cosmology of the KKLT
models is challenging.  For in these cases, there is always another
modulus in the hidden sector. For the realistic cosmological scenarios,
we need to include those fields in the discussion. We will leave more
detailed discussion of the cosmology for another publication, but we
noted that it is likely that these moduli are very heavy, and their
decays may reheat the universe to temperatures above those of
nucleosynthesis.  With the addition of vector-like fields to the MSSM,
these decays can lead to an acceptable dark matter density.  We conclude
from these observations that the idea that decays of heavy moduli heat
the universe above nucleosynthesis temperatures, producing the dark
matter in their decays, remains a viable one; the reader can judge
whether it is more plausible than scenarios without light moduli.

The possibility that inflaton decays to gravitinos are not helicity
suppressed raises the prospect of significant constraints on
inflationary cosmology.  Here, however, we again found that there can be
significant suppression of inflaton decays to gravitinos.  The results
depend on the structure of the hidden sector. 
Conditions for the suppression are the same as those in the discussion
of the moduli decays.
For example, if we take the O'Raifeartaigh model for supersymmetry
breaking, the chiral superfield $z$, which gets $F$-term, is charged
under $R$-symmetry which is broken only by the small constant term in
the superpotential. We may naively expect absence or a small coefficient
in the $\phi^\dagger z z$ term in the \Kahler potential in this
case. The supersymmetry breaking mass term for $z$ appears through the
one-loop effect, which can be significantly larger than the gravitino
mass.
In the case where the inflaton is heavier than the scalar field $z$, the amplitude of the
inflaton decay into two gravitinos is suppressed by $(m_{\rm
Hidden}/m_\phi)^2$ or $m_{3/2}/m_\phi$.  This is sufficient to avoid the
constraints from nucleosynthesis or overproduction of the dark matter in
many inflation models. Problems with the decay of the hidden sector
particle can be avoided for low enough reheating temperature.  So it
appears that the constraints on inflation from inflaton decays to
gravitinos are quite mild.

\noindent {\bf Acknowledgements:}

\noindent 
We thank T.~Banks, M.~Peskin and especially Y.~Shadmi for
conversations.  The work of M.D. and 
A.M. was supported in part by the U.S. Department of Energy. M.D.
and Y.S. acknowledge support from the U.S.-Israel Binational Science
Foundation and thank the Technion for its
hospitality during the completion of this work. The work of R.K. was
supported by the U.S. Department of Energy under contract number
DE-AC02-76SF00515.  
The work of Y.S. is supported by the US Department of Energy
under contract W-7405-ENG-36.

\end{document}